\documentclass[
  aps,
  prl,
  reprint,
  superscriptaddress,
  amsfonts,amssymb,amsmath,
]{revtex4-2}

\usepackage{bbm}
\usepackage{amsmath, amssymb, amsfonts, latexsym, mathtools}
\usepackage{verbatim}
\usepackage{epsfig}
\usepackage{physics}
\usepackage{appendix}
\usepackage{bm}
\usepackage{times}
\usepackage{graphicx}
\usepackage{hyperref}
\hypersetup{
  setpagesize  = false,
  colorlinks   = true,
  urlcolor     = blue,
  linkcolor    = blue,
  citecolor    = blue
}

\newcommand{\Pf}{\text{Pf}}
\newcommand{\matidentity}{\mathbb{I}}
\newcommand{\HC}{\text{H.c.}}

\newcommand{\beginsupplement}{
  \setcounter{equation}{0}
  \renewcommand{\theequation}{S\arabic{equation}}%
  \setcounter{figure}{0}
  \renewcommand{\thefigure}{S\arabic{figure}}%
  \setcounter{table}{0}
  \renewcommand{\thetable}{S\arabic{table}}%
  \setcounter{section}{0}
  \renewcommand{\thesection}{S\Roman{section}}%
  \makeatletter
    \renewcommand{\theHequation}{S\arabic{equation}}%
  \makeatother
}

\begin{document}

%%% title %%%
\title{Entanglement negativity in free fermions: twisted characteristic polynomial, \\ universal bounds, and area laws}

%%% author(s) %%%
\author{Ryota Matsuda}
\email{r-matsuda918@g.ecc.u-tokyo.ac.jp}
\affiliation{Department of Physics, The University of Tokyo, 7-3-1 Hongo, Bunkyo-ku, Tokyo 113-0033, Japan}
\author{Zongping Gong}
\affiliation{Department of Applied Physics, The University of Tokyo, 7-3-1 Hongo, Bunkyo-ku, Tokyo 113-8656, Japan}

%%% date %%% 
\date{\today}

%%% abstract %%%
\begin{abstract}
We present a general and simple formula for computing the entanglement negativity in free fermions. Our formula allows for deriving several universal bounds on negativity and its rate of change in dynamics. The bound on negativity directly relates the clustering property of correlations in free-fermion states to the entanglement area law, and provides the optimal condition for the area law in mixed free fermion states with long-range correlations. In addition, we establish an area-law bound on entanglement generation in open systems, analogous to previously known results for entanglement entropy in unitary dynamics. Our work provides new analytical insights into fermionic mixed-state entanglement.
\end{abstract}

\maketitle

%%%========== main text begin ==========%%%
\textit{Introduction.---} Understanding quantum entanglement in mixed states is a central challenge in modern physics and quantum information science. This is essential for deciphering the quantum features in realistic systems at finite temperatures or coupled to an environment~\cite{Horodecki-QuantumEntanglement-2009d,Horodecki-Mixed-StateEntanglement-1998r,Wootters-EntanglementFormationQubits-1998w,Peres-SeparabilityCriterionMatrices-1996j}. Entanglement negativity has emerged as an indispensable tool for this challenge, providing a computable measure that successfully captures quantum correlations beyond the scope of entanglement entropy for bipartite pure states~\cite{Vidal-ComputableMeasureEntanglement-2002c}. Its application to both (multipartite) closed and open quantum many-body systems has yielded fruitful insights across various contexts, %settings, 
including critical phenomena~\cite{Calabrese-EntanglementNegativityTheory-2012t,Calabrese-FiniteTemperatureTheory-2015v,Sherman-Nonzero-temperatureEntanglementDeath-2016k}, nonequilibrium dynamics~\cite{Wichterich-ScalingEntanglementCriticality-2009y,Eisler-EntanglementNegativityEquilibrium-2014y}, and topological phases~\cite{Lee-EntanglementNegativityOrder-2013i,Castelnovo-NegativityTopologicalCode-2013s}. 
In particular, its computability has served as a powerful analytical tool for deriving rigorous results about mixed-state entanglement in bosonic and spin systems~\cite{Eisert-Noise-drivenQuantumCriticality-2010j,Cramer-Entanglement-areaLawSystems-2006q,Cramer-CorrelationsSpectralLattices-2006p,Eisler-EntanglementNegativityEquilibrium-2014y,Sherman-Nonzero-temperatureEntanglementDeath-2016k,Audenaert-EntanglementPropertiesChain-2002a}.

However, extending the notion of negativity to fermionic systems turns out to be highly nontrivial, as the entanglement structure differs fundamentally from that of bosons due to the anticommuting nature of fermionic operators and the intrinsic superselection rules~\cite{Wick-IntrinsicParityParticles-1952z,Banuls-EntanglementFermionicSystems-2007n,Spee-ModeEntanglementStates-2018d,Verstraete-QuantumNonlocalityProtocols-2003e,Schuch-NonlocalResourcesRules-2004m,Schuch-QuantumEntanglementRules-2004y}. 
These properties necessitate a different definition of negativity to correctly capture fermionic entanglement. %and such 
A breakthrough was made in Ref.~\cite{Shapourian-PartialTime-reversalSystems-2017q}, %. This definition comes from reinterpreting 
which reinterprets the partial transpose as a partial time-reversal transformation. %, which 
This allows for a quantum-information-theoretically consistent extension to fermionic systems~\cite{Shapourian-PartialTime-reversalSystems-2017q,Shiozaki-Many-bodyTopologicalSymmetries-2018j}.
%the partial transpose can be  should further be twisted by the subsystem fermion-parity operator. 
%Such an additional twist accounts for the anticommuting nature and restores the Hermiticity of the partially transposed density operator. 
%However, since 
The transposed density matrix can be restored to be Hermitian by the fermion parity operator~\cite{Shapourian-TwistedUntwistedFermions-2019u}. Nevertheless, %is no longer Hermitian, 
both analytical and practical computations of negativity turn out to be extremely involved even for free fermions. %, despite the conceptual simplicity. 
It thus remains elusive to rigorously analyze universal properties of fermionic mixed-state entanglement.

In this Letter, we demonstrate that the negativity in free fermions
is simply determined by the zeros of a ``twisted'' characteristic polynomial of the covariance matrix. As shown in Eq.~(\ref{tcp}), the twist is realized by replacing the variable in one diagonal block with its minus inverse. 
Utilizing our formula and the monotonicity of negativity under local operations, we establish upper and lower bounds on both the negativity and its rate of change in dissipative dynamics \footnote{Prior work exists~\cite{Eisert-EntanglementNegativityStates-2018u} deriving rigorous upper and lower bounds on negativity using Gaussian channels, but these bounds concern the naive bosonic negativity, which cannot fully capture fermionic entanglement such as that of Majorana bonds~\cite{Shapourian-PartialTime-reversalSystems-2017q}.}. 
By combining these results with locality, we discuss entanglement area laws in static and dynamical settings. 
%In particular, 
Regarding the static case, we unveil how entanglement is related to the clustering property of the system. Notably, application of this argument to the Gibbs states answers an open problem in Ref.~\cite{Kim-ThermalAreaSystems-2025u} concerning long-range systems.
In addition, the dynamical area law represents the first attempt to extend previous rigorous results about %dynamical area laws in 
unitary dynamics~\cite{VanAcoleyen-EntanglementRatesLaws-2013v,Bravyi-Lieb-RobinsonBoundsOrder-2006h,Bravyi-UpperBoundsHamiltonians-2007b,Gong-EntanglementAreaSystems-2017e} to nonunitary settings.

\textit{Explicit formula of the negativity for free fermions.---} 
We consider a general fermionic system with $N$ modes and define the fermionic creation and annihilation operators $\hat{f}_j^\dagger,\hat{f}_j$ for each mode. From these operators we can construct $2N$ Majorana operators $\hat c_j$ ($j=1,2,...,2N$) by $\hat{c}_{2j-1} = \hat{f}_j +\hat{f}_j^\dagger\;,\;\hat{c}_{2j} = -i(\hat{f}_j-\hat{f}_j^\dagger)$ satisfying $\{\hat{c}_j,\hat{c}_{j'}\} = 2\delta_{jj'}$.
Any free-fermion state $\hat{\rho}$ is given by~\cite{Bravyi-LagrangianRepresentationOptics-2004s}
\begin{align}
    \hat{\rho} = Z^{-1} e^{\frac{1}{4}\sum_{j,j'}\Omega _{j,j'}\hat{c}_j \hat{c}_{j'}},
\end{align}
where $\Omega$ is a $2N\times 2N$ purely imaginary antisymmetric matrix and $Z$ is a normalization constant such that $\Tr \hat{\rho}=1$. This is a Gaussian state fully characterized by its covariance matrix
\begin{equation}
\Gamma_{jj'} = \frac{1}{2}\Tr[[\hat{c}_j,\hat{c}_{j'}]\hat{\rho}].
\end{equation}
Since $\Gamma$ is related to $\Omega$ via $\Gamma = \tanh \frac{\Omega}{2}$, its spectrum consists of pairs $\pm \nu_j\in [-1,1]$ for $j=1,2,\ldots,N$. 
To study bipartite entanglement, we partition the covariance matrix into blocks %matrices 
corresponding to subsystems $A$ with $N_A$ modes and $B$ with $N_B=N-N_A$ modes:
\begin{align}
    \Gamma = \begin{pmatrix}
        \Gamma_{A} & \Gamma_{AB}\\
        \Gamma_{BA} & \Gamma_{B}
    \end{pmatrix}. \label{cov_partition}
\end{align}
The logarithmic entanglement negativity with respect to such a bipartition is given by \cite{Vidal-ComputableMeasureEntanglement-2002c}
\begin{equation}
\mathcal{E}=\ln\|\hat\rho^{{\rm T}_A}\|_1,
\label{ENdef}
\end{equation}
where ${\rm T}_A$ denotes the partial time-reversal transformation %transpose 
of subsystem $A$ and $\|O\|_1={\rm tr}\sqrt{O^\dagger O}$ denotes the trace norm (i.e., the sum of singular values of $O$). %As we noted, the partial transpose is equivalent to the partial time-reversal transformation, which is different from the naive definition of partial transpose in fermions.
Since $\hat \rho^{{\rm T}_A}$ is no longer Hermitian, %in fermions
we add a parity %additional 
twist to make it Hermitian without changing the trace norm \cite{Shapourian-TwistedUntwistedFermions-2019u}:
\begin{align}
    \hat\rho^{\tilde{{\rm T}}_A} = \hat \rho^{{\rm T}_A}\hat P_A,
    \label{tpt}
\end{align}
where $\hat P_A=\prod^{N_A}_{j=1}(i\hat c_{2j-1}\hat c_{2j})$ is the subsystem fermion parity operator \cite{Fidkowski-TopologicalPhasesDimension-2011b}. The negativity (\ref{ENdef}) is then computable from the (real) spectrum of $\hat\rho^{\tilde{\rm T}_A}$. %Since this twisted partial transpose $\hat\rho^{\tilde{{\rm T}}_A}$ and its corresponding covariance matrix are Hermitian, negativity is computed simply by its spectrum. 
%We clarify the action of this twisted partial transpose on the level of characteristic polynomial for the covariance matrix, which determines its spectrum.%, thereby giving the following formula of negativity.

%To compute negativity Eq.~(\ref{ENdef}) in terms of the covariance matrix Eq.~(\ref{cov_partition}), %fermionic entanglement negativity, 
%Specifically, 
Concerning the effect of this twisted partial transpose (\ref{tpt}) on the level of covariance matrix $\Gamma$, we define the twisted characteristic polynomial %of $\Gamma$ 
as 
\begin{align}
    P(\lambda) &= \lambda^{2N_A} \det\begin{pmatrix}
    \Gamma_A+ \lambda^{-1} \matidentity_A & \Gamma_{AB} \\
    \Gamma_{BA} & \Gamma_B -\lambda \matidentity_B
    \end{pmatrix}.\label{tcp}
\end{align}
Note that $P(\lambda)$ is different from the ordinary characteristic polynomial by the transformation $\lambda \to -\lambda^{-1}$ in the subsystem $A$. From $P(\lambda)$, we define $P_>(\lambda)$ as its factor %this polynomial 
with zeros $\lambda_j$ having absolute value larger than 1:
\begin{align}
    P_>(\lambda) = \frac{P(\lambda)}{\prod_{j:|\lambda_j|\leq1}(\lambda-\lambda_j)}.\label{poly_>_def}
\end{align}
%Then, 
We find the logarithmic negativity (\ref{ENdef}) simply reads %is simply expressed as
\begin{align}
    \mathcal{E} = \frac{1}{2} %\qty
    \ln |P_>(0)| .\label{negativity_formula}
\end{align}

We note that our formula is consistent with the results in %the direct consequence of the method based on the covariance matrix of the twisted partial transpose in 
Ref.~\cite{Shapourian-TwistedUntwistedFermions-2019u} %. Indeed, 
%under the assumption of 
provided the invertibility of $\Gamma_A$%, both methods are equivalent
~\cite{Supplementary_material}. 
%the zeros of twisted characteristic polynomial exactly yields the eigenvalues of the covariance matrix of the twisted partial transpose proposed in Ref.~\cite{Shapourian-TwistedUntwistedFermions-2019u}, under the assumption of the invertibility of $\Gamma_A$~\cite{Supplementary_material}.
%In this sense, both methods are equivalent. 
Nonetheless, our formula is applicable regardless of the invertibility, %of $\Gamma_A$, 
and this is necessary to derive the following universal bounds.

\textit{Bounds on the negativity.---} We show that our formula \eqref{negativity_formula} is useful for deriving upper and lower bounds on negativity when combined with the monotonicity of negativity under local operations~\cite{Shapourian-EntanglementNegativityStates-2019q}. A key observation is %To motivate our approach, we note 
that if $\Gamma_A$ or $\Gamma_B$ vanishes, Eq.~\eqref{negativity_formula} reduces to a remarkably simple formula
\begin{align}
    \mathcal{E} = \frac{1}{2}\tr[\ln(\matidentity_B+\Gamma_{BA}\Gamma_{AB})].
\end{align}
This means that if one applies local operations to transform some $\Gamma$ with vanishing %so that 
$\Gamma_A$ or $\Gamma_B$ into / from the target state, %vanishes, 
negativity can be bounded using the off-diagonal blocks of $\Gamma$, %the output covariance matrix, 
which is responsible for the inter-subsystem correlation.

Concretely, a general Gaussian operation on a fermionic Gaussian state with covariance matrix $\Gamma_{\rm in}$ is given by~\cite{Bravyi-LagrangianRepresentationOptics-2004s}
\begin{align}
    \Gamma_{\text{out}} = K(\Gamma_{\text{in}}^{-1}+D)^{-1}K^\dagger +C, \label{gaussian_operation}
\end{align}
where $\Gamma_{\rm out}$ denotes the covariance matrix of the output state, $C,D,K$ are all purely imaginary $2N\times 2N$ matrices, and $C,D$ are antisymmetric. %Also, 
There is an additional constraint
\begin{align}
    -\matidentity \oplus \matidentity \leq \begin{pmatrix}
            C & K\\
            K^\dagger & D
        \end{pmatrix} \leq \matidentity \oplus \matidentity \label{CP_constraint}
\end{align}
to ensure the operation is completely positive. Here $\matidentity = \matidentity_A\oplus \matidentity_B$ is the global identity. To make such an operation (\ref{gaussian_operation}) local, we further require $C, D, K$ to be block diagonalized. %By 
Taking $D=0,K\propto \matidentity, C \propto \Gamma_A \oplus \Gamma_B$ and defining $k_{\pm} = 1 \pm \max \{\|\Gamma_{A}\|,\|\Gamma_{B}\|\}$, where $\|\cdot\|$ denotes the operator norm %, i.e., the 
(largest singular value), we %can derive 
obtain an upper bound~\cite{Supplementary_material}
\begin{align}
    \mathcal{E} \leq \frac{1}{2}\tr[\ln(\matidentity_B+k_-^{-2}\Gamma_{BA}\Gamma_{AB})]
    \label{upperbound}
\end{align}
%with constraint 
conditioned on $\|\Gamma_{AB}\|\leq k_-$ for $\Gamma_{\rm out}=\Gamma$. Also, for $\Gamma_{\rm in}=\Gamma$, there is always a lower bound
\begin{align}
    \mathcal{E} \geq \frac{1}{2}\tr[\ln(\matidentity_B+k_+^{-2}\Gamma_{BA}\Gamma_{AB})].\label{lowerbound}
\end{align}
For later proofs, we use a looser but simpler version of these bounds: 
\begin{align}
    \frac{\ln\qty(1+k_+^{-2} \|\Gamma_{AB}\|^2)}{2\|\Gamma_{AB}\|^2}\|\Gamma_{AB}\|_2^2 \leq \mathcal{E} \leq\frac{1}{2} k_{-}^{-2} \|\Gamma_{AB}\|^2_2, \label{simplerbounds}
\end{align}
where $\|O\|_2=\sqrt{{\rm tr}(O^\dagger O)}$ denotes the Frobenius norm. 

\begin{figure}[t]
	\centering
	\includegraphics[width=1.0\linewidth]{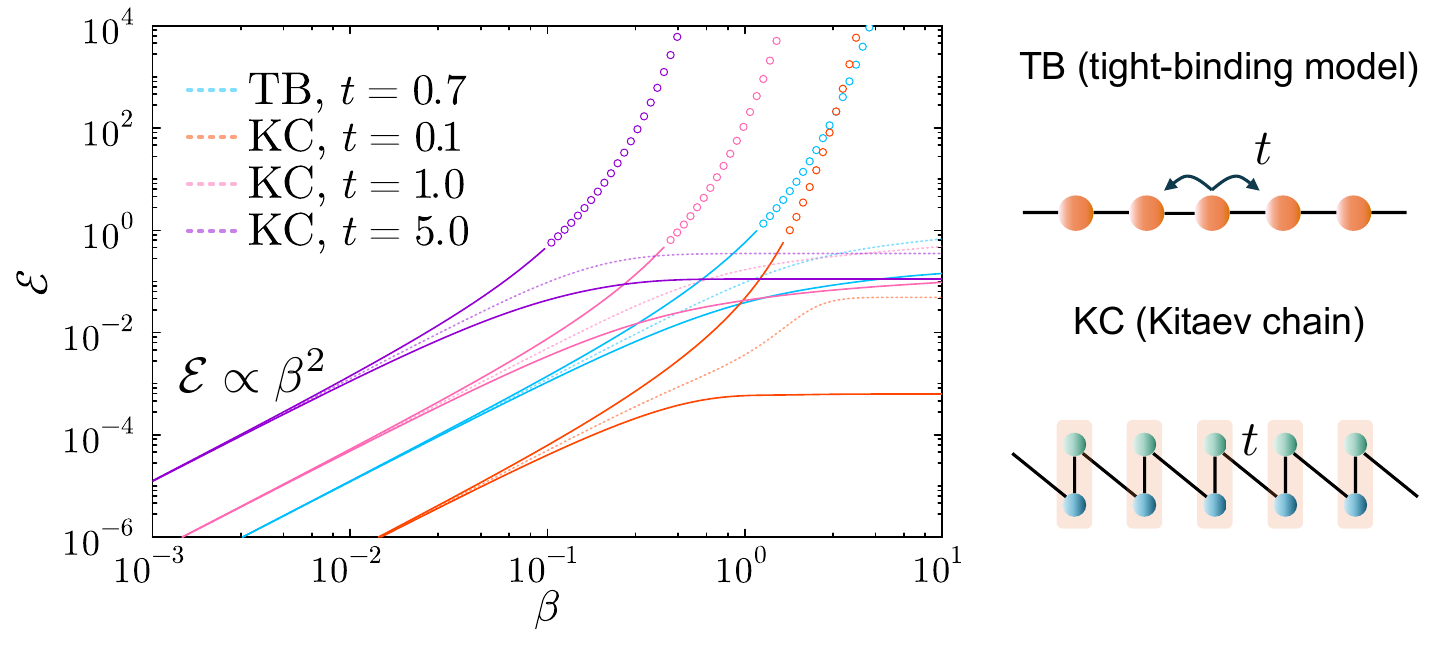}
	\caption{Universal bounds \eqref{simplerbounds} on the negativity of the Gibbs states. The bounds (exact value) are shown in the solid (dotted) curve, with the circle marker denoting the regime $\|\Gamma_{AB}\| > k_-$, where the upper bound is %unjustified.
    not applicable.
    %TB corresponds to the tight-binding chain with hopping constant $t=0.7$, while $t=0.1,1.0,5.0$ shows the hopping constant of the Kitaev chain (see SM~\cite{Supplementary_material} for the precise definition of the model). 
    TB and KC correspond to the tight-binding chain and the Kitaev chain \cite{YuKitaev-UnpairedMajoranaQuantumwires-2001z}, respectively.
    In the Kitaev chain, $t=1$ is the critical value, and $t>1(<1)$ corresponds to the topological (trivial) phase at zero temperature. In both models, we take $N=20$ and adopt open boundary conditions. Also, we set the subsystem $A$ as half of the chain.}
	\label{fig:bound_high_temp}
\end{figure}

We numerically demonstrate these bounds \eqref{simplerbounds} using the Gibbs states of two different free-fermion models in Fig. \ref{fig:bound_high_temp}. In both cases, our bound is tight in the high temperature regime and scales as $\beta^2$ \footnote{This universal scaling was derived in~\cite{Choi-Finite-temperatureEntanglementDimension-2024f} using high-temperature expansion technique.}. %, regardless of the choice of models. [ZG: ``in both cases'' already means ``regardless of the choice of models''] 
This universal scaling follows from the factor $\|\Gamma_{AB}\|^2_2$ in our bounds \eqref{simplerbounds}, because the high-temperature asymptotic form of the covariance matrix %in the high temperature limit 
reads $\Gamma \simeq -2\beta H$, with $H$ being the matrix specifying the quadratic Hamiltonian $\hat{H} = \sum_{j,j'} H_{j,j'} \hat{c}_j\hat{c}_{j'}$.
This behavior is in stark contrast with bosons and spins, whose %where 
entanglement completely vanishes at some finitely high temperature~\cite{Arnesen-NaturalThermalModel-2001f,Anders-EntanglementSeparabilityTemperature-2008f,Bakshi-High-TemperatureGibbsPreparable-2024t}. This qualitative difference arises from %manifests itself in 
the fact that fermions are much more difficult to be unentangled under the superselection rule \cite{Shapourian-EntanglementNegativityStates-2019q,Akshar-High-TemperatureFermionicStates-2025s,Parez-FateEntanglement-2024m}. Note that a similar reasoning can %be used to 
rule out the possibility of such entanglement ``sudden death'' %of entanglement 
also in dissipative dynamics~\cite{Yu-Finite-timeDisentanglementEmission-2004q,Caceffo-FateEntanglementSystems-2024j,Parez-FateEntanglement-2024m,Gong-QuantumMany-BodyTime-2024e}, which is the next setting 
we consider.

\textit{Bound on the negativity change rate.---} %As a second setup, We move on to 
Let us consider a time-evolving free-fermion state %specified by 
$\Gamma(t)$ and derive a bound on the negativity change rate $\partial_t \mathcal{E}(t)$. %This is achievable by 
To this end, we first figure out the explicit formula for $\partial_t \mathcal{E}(t)$ %the negativity change rate 
from Eq.~\eqref{negativity_formula} as~\cite{Supplementary_material}\footnote{Although we will not use this fact in the subsequent proofs or discussions, it is worth noting that an integral representation of entanglement negativity can be derived by artificially setting $\Gamma(t)=t\Gamma$ for $t\in[0,1]$ in this formula.}
\begin{align}
    \pdv{\mathcal{E}(t)}{t} =  \frac{1}{2}\Tr[\mathcal{P}_{AB}(\Gamma(t))\pdv{\Gamma(t)}{t}], \label{negativity_rate_formula}
\end{align}
where $\mathcal{P}_{AB}$ is a superoperator defined by
\begin{align}
    \mathcal{P}_{AB}(\Gamma) = \int_{-\pi}^{\pi} \frac{dk}{2\pi} (Z_{AB}(k)+\Gamma)^{-1}\label{PAB_def},
\end{align}
with $Z_{AB}(k) = e^{ik} \matidentity_A\oplus(-e^{-ik})\matidentity_B$ exhibiting a twist structure inheriting from Eq.~(\ref{tcp}). Note that $\partial_t\mathcal{E}(t)$ may become discontinuous at some specific time, when $\mathcal{P}_{AB}(\Gamma(t))$ becomes ill-defined. Nevertheless, we can show that $\|\mathcal{P}_{AB}(\Gamma)\|\le 2$ whenever defined. %is $\mathcal{O}(1)$ with respect to the system size. 
By applying H\"older's inequality, we can bound this change rate as~\cite{Supplementary_material}
\begin{align}
    \qty|\pdv{\mathcal{E}(t)}{t}| \leq \left\|\pdv{\Gamma(t)}{t} \right\|_1.
    \label{negativity_rate_bound}
\end{align}
%where $\|O\|_1={\rm tr}\sqrt{O^\dagger O}$ denotes the trace norm. 
Intuitively, this bound means %is quite natural: 
the speed of entanglement change never exceeds that of the covariance matrix, which includes both intra- and inter-subsystem correlations. 

Let us apply %this bound 
Eq.~\eqref{negativity_rate_bound} to the Lindblad dynamics \cite{Gorini-PositiveDynamicalSystems-1976x,Lindblad-GeneratorsQuantumSemigroups-1976s}
\begin{align}
    \pdv{\hat{\rho}}{t} = -i[\hat{H},\hat{\rho}] + \sum_\mu\qty(\hat{L}_\mu\hat{\rho}\hat{L}_\mu^\dagger - \frac{1}{2}\{\hat{L}_\mu^\dagger \hat{L}_\mu, \hat{\rho}\}), \label{lindblad_dynamics}
\end{align}
where $\hat{L}_\mu$ are the Lindblad operators arising from dissipation. We assume that the Lindblad operators are linear in the Majorana operators, so that the time-evolved states remain Gaussian~\cite{Barthel-SolvingQuasi-freeSystems-2022u,Bardyn-TopologyDissipation-2013j,Eisert-Noise-drivenQuantumCriticality-2010j}. 
%Since we consider Gaussian dynamics, 
Accordingly, the time-evolution of the states \eqref{lindblad_dynamics} is equivalently described by the equation of motion for the covariance matrix~\cite{Barthel-SolvingQuasi-freeSystems-2022u,Bardyn-TopologyDissipation-2013j,Eisert-Noise-drivenQuantumCriticality-2010j}:
\begin{align}
    \pdv{\Gamma}{t} = -i[4H,\Gamma]-\{X,\Gamma\}+2Y, \label{cov_dynamics}
\end{align}
with $H$ being the Hamiltonian matrix, and $X=2B_R$, $Y=2iB_I$. %Note that 
Here $B_R$ and $B_I$ are the real and imaginary part of the matrix $B_{ij} = \sum_\mu L_{\mu,i}^* L_{\mu,j}$ determined %defined 
from the Lindblad operators %as 
$\hat{L}_\mu = \sum_j L_{\mu,j}\hat{c}_j$. Combining Eq.~\eqref{cov_dynamics} with Eq.~\eqref{negativity_rate_bound}, we arrive at
\begin{align}
    \qty|\pdv{\mathcal{E}(t)}{t}| \leq 2(4\left\|H\right\|_1+\|X\|_1+\|Y\|_1),
\end{align}
where the right-hand side depends only on the Lindbladian. %which bounds the absolute value of the negativity change rate using the generators of Lindblad dynamics.

This bound can be viewed as an open-system analogue of the small incremental entangling (SIE) theorem for unitary dynamics \cite{Bravyi-UpperBoundsHamiltonians-2007b,VanAcoleyen-EntanglementRatesLaws-2013v}. %Therein, 
It claims the absolute rate of change of entanglement entropy is upper bounded by $c\|\hat H_{AB}\|\ln d$, where $c$ is some $\mathcal{O}(1)$ constant, $\hat H_{AB}$ is the coupling Hamiltonian between two subsystems $A,B$, and $d=\min (d_A,d_B)$ with $d_{A(B)}$ denoting the local Hilbert-space dimension of %subsystem 
$A(B)$ %on which $H_{AB}$ acts nontrivially 
\footnote{Although the name of SIE theorem suggests that it concerns the entanglement generation, the theorem also applies to the entanglement destruction}. Note that $d_{A(B)}=2^{N_{A(B)}}$ and $\| \hat H\|=\|H\|_1$ for $\hat H=\sum_{j,j'} H_{j,j'}\hat c_j\hat c_{j'}$, so the results are indeed comparable except for an additional $\min(N_A,N_B)$ in the SIE bound.

%This observation leads us to 
We can even derive a tighter bound concerning the increase rate of entanglement by extracting the contribution from inter-subsystem hopping, pairing and dissipation. 
To this end, we utilize the fact that negativity rate~\eqref{negativity_rate_formula} is linear with respect to $\partial_t\Gamma(t)$, which is in turn linear in the generators $H, X, Y$. Therefore, the negativity dynamics can be %rigorously 
decomposed into contributions from intra-subsystem parts corresponding to local operations (LO) and %one that generates 
inter-subsystem parts: %correlations:
\begin{align}
    \pdv{\mathcal{E}}{t} =  \left.\pdv{\mathcal{E}}{t}\right|_{\text{LO}} +  \left.\pdv{\mathcal{E}}{t}\right|_{\text{inter}}.
\end{align}
%This decomposition is possible because the negativity rate \eqref{negativity_rate_formula} is linear with respect to $\partial_t\Gamma(t)$, and the equation of motion for the covariance matrix \eqref{cov_dynamics} is also linear. 
Specifically, $\left.\partial_t\mathcal{E}(t)\right|_{\text{LO}}$ and $\left.\partial_t\mathcal{E}(t)\right|_{\text{inter}}$ are given by Eq.~\eqref{negativity_rate_formula} with $H,X,Y$ in Eq.~\eqref{cov_dynamics} replaced by their block-diagonal terms $H_0, X_0, Y_0$ and off-diagonal terms $H_{\text{inter}}, X_{\text{inter}}, Y_{\text{inter}}$, respectively.
Since Local operations never increase the entanglement between $A$ and $B$ \footnote{We give a proof of this fact in SM}, the change rate of negativity is upper bounded by %the contribution from the inter-correlation part:
\begin{align}
    \pdv{\mathcal{E}}{t} \leq \left. \pdv{\mathcal{E}}{t} \right|_{\text{inter}}.
\end{align}
Applying %the bound 
Eq.~\eqref{negativity_rate_bound} to this inter-correlation term then yields
\begin{align}
    \pdv{\mathcal{E}}{t} \leq 2 (4\|H_{\text{inter}}\|_1 + \|X_{\text{inter}}\|_1 + \|Y_{\text{inter}}\|_1).\label{negativity_increase_rate_bound}
\end{align}
%Since $\|H_{\text{inter}}\|_1\ \le 4\|H_{AB}\| \log_2d$, our bound for the increase rate Eq.~\eqref{negativity_increase_rate_bound} reproduces the same scaling as the SIE theorem in unitary dynamics.

\textit{Clustering property and area law of negativity.---} Let us return to the static setup and discuss the entanglement area law of lattice systems with locality. Recently, significant progress has been made for extending the entanglement area law to systems with long-range interactions decaying by power-law \cite{Kuwahara-AreaLawSystems-2020q,Kim-ThermalAreaSystems-2025u,Gong-EntanglementAreaSystems-2017e,Vodola-Long-rangeIsingModes-2015g}. One notable feature of such systems is that the correlation functions exhibit the power-law decay at high temperatures %even in the non-critical phase, inheriting the locality of the system 
\cite{Kim-ThermalAreaSystems-2025u,Vodola-Long-rangeIsingModes-2015g}. %Therefore, the optimal condition 
A fundamental question is to identify the minimal decay exponent for validating the area law. %in such long-range systems. %is of fundamental interest. 
We approach this problem by using %our bounds shown in 
Eq.~\eqref{simplerbounds}, which directly relates the clustering property and area law.

Suppose the system lives on a $D$-dimensional lattice $\Lambda$, and we have $j=(\boldsymbol{r},s)$ with $\boldsymbol{r}\in\Lambda$ denoting a lattice site and $s\in I$ labeling an internal degree of freedom. %To include 
For long-range systems, 
we consider the following weak notion of clustering:
\begin{align}
    \|\Pi_{\bm{r}}\Gamma\Pi_{\bm{r}'}\| \leq CF_\alpha(|\bm{r}-\bm{r}'|),
    \label{clustering}
\end{align}
where $\Pi_{\boldsymbol{r}}$ is the projector onto site $\boldsymbol{r}$, $C$ is an $\mathcal{O}(1)$ constant, and $F_\alpha(r)=(r+1)^{-\alpha}$ expresses power-law decay with its exponent $\alpha$. The key observation from our bound \eqref{simplerbounds} is that the area law holds if and only if $\|\Gamma_{AB}\|_2^2=\sum_{\bm{r}\in A, \bm{r}'\in B} \|\Pi_{\bm{r}}\Gamma\Pi_{\bm{r}'}\|_2^2$ is proportional to its boundary $|\partial A|$, since  $k_\pm$ and $\|\Gamma_{AB}\|$ are $\mathcal{O}(1)$ quantities with respect to the subsystem size. In $D$ dimensions, we can evaluate the summation using $g(r) = \int_r^\infty dR F(R)^2 R^{D-1}$ and $G(r) = \int_r^\infty dR g(R)$ as~\cite{Gong-EntanglementAreaSystems-2017e} 
\begin{align}
    \mathcal{E} \leq c_gC^2|I||\partial A|G(\text{dist}(A,B)), \label{area_law_eq_state}
\end{align}
where $c_g$ is an $\mathcal{O}(1)$ constant depending on the lattice geometry, and $\text{dist}(A,B) = \min\{|\bm{r}-\bm{r}'|:\bm{r}\in A,\bm{r}'\in B\}$ is the distance between subsystems $A$ and $B$. The area law follows %is regarded 
as a special case with $B=\bar{A}$ and %replacing 
$\text{dist}(A,B)=1$. %by 1. 
Therefore, we %derive 
identify the area-law condition %for the area law 
as $\alpha > (D+1)/2$, so that $G(1)$ is convergent.
While this argument relies on the upper bound \eqref{upperbound} conditioned on %which is applicable under the condition 
$\|\Gamma_{AB}\| \le k_{-}$, %it 
the condition is easily justified when the state is sufficiently mixed, %distant from pure states: 
i.e., the purity $\tr[\hat\rho^2]$ is small enough.

Our result can be examined %confirmed 
in Gibbs states $\hat{\rho}=e^{-\beta\hat{H}}/Z$ with %its locality allowing 
$\hat H$ involving long-range hopping and pairing:
\begin{align}
    \|\Pi_{\bm{r}}H\Pi_{\bm{r}'}\| \leq hF_\alpha(\|\bm{r}-\bm{r}'\|),\label{locality}
\end{align}
%Since free fermions do not exhibit thermal phase transitions at finite temperature, 
with $\mathcal{O}(1)$ constant $h$. For $\alpha>D$, one can prove that the clustering property (\ref{clustering}) follows from Eq.~(\ref{locality}). %is naturally expected. 
%Indeed, 
While a rigorous proof is absent in the regime $\alpha \le D$, the clustering property is numerically verified in Ref.~\cite{Kim-ThermalAreaSystems-2025u} for various %specific 
models. %, regardless of the power $\alpha$. 
Since the condition for the upper bound $\|\Gamma_{AB}\| \le k_{-}$ is easily satisfied in the high temperature regime, our argument concludes the condition for this thermal area law as $\alpha > (D+1)/2$, which confirms the conjecture %exactly coincides 
%with numerical results 
in Ref.~\cite{Kim-ThermalAreaSystems-2025u} based on numerical observations. 

We remark that our argument only relies upon the clustering property. This implies, for example, that our argument could be extended to the nonequilibrium steady state of long-range Lindblad dynamics \cite{Gorini-PositiveDynamicalSystems-1976x,Lindblad-GeneratorsQuantumSemigroups-1976s,Passarelli-DissipativeTimeLindbladians-2022t,deAlbornoz-EntanglementTransitionLindbladians-2024q}.

\textit{Area law of negativity increase rate.---} Finally, we establish a dynamical area law for the \emph{increase} rate of negativity in %local 
Lindblad dynamics by utilizing our bound \eqref{negativity_increase_rate_bound}. 
More precisely, %dissipative dynamics. 
we will prove
\begin{align}
    \pdv{\mathcal{E}}{t} \leq \mathcal{O}(|\partial A|),\label{negativity_rate_area_law_statement}
\end{align}
for the Lindblad dynamics \eqref{lindblad_dynamics} subject to locality constraints. Unlike the previous section, we do not assume the Gaussian state satisfies clustering or any other specific properties.  

\begin{figure}[t]
    \centering
    \includegraphics[width=\linewidth]{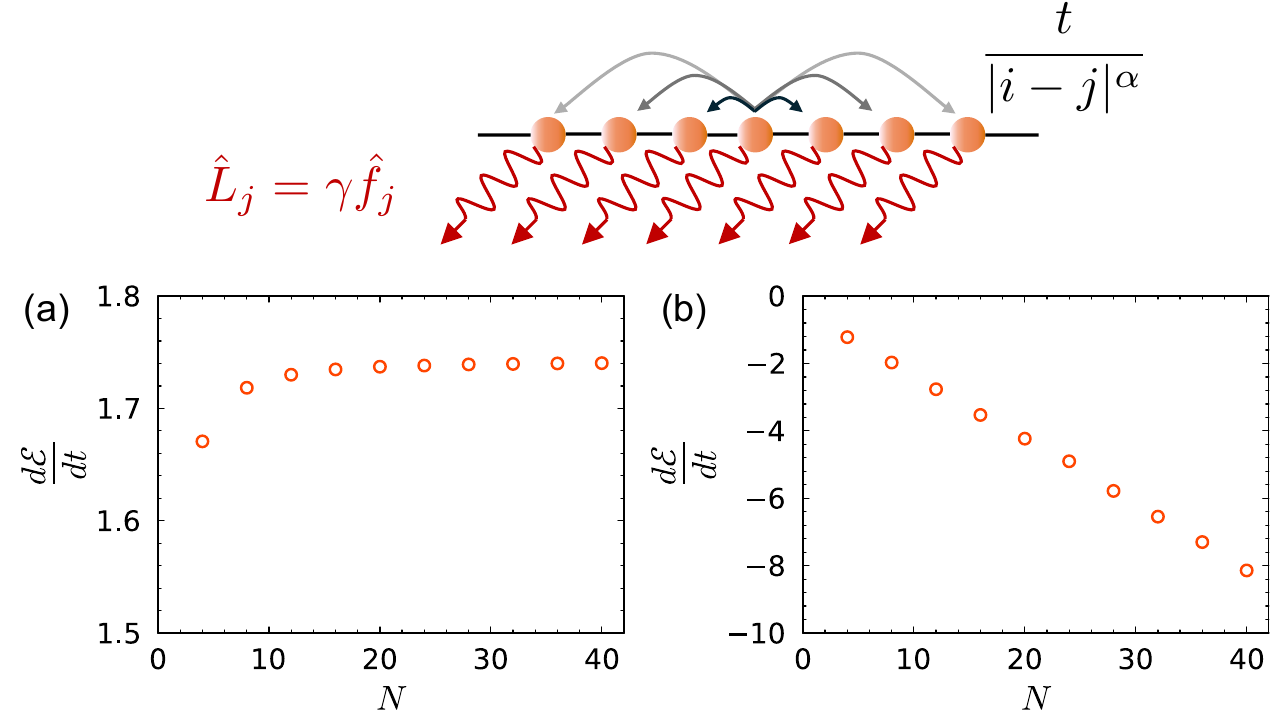}
    \caption{%The 
    Change rate of negativity 
     in a dissipative long-range hopping model %$\|\Pi_{\bm{r}}H\Pi_{\bm{r}'}\| = {1}/{|\bm{r}-\bm{r}'|^{2.1}}$ 
     with $t = 1, \alpha=2.1$ and particle loss %at a 
     rate of $\gamma=0.5$. The system is one-dimensional with open boundary conditions, and the subsystem $A$ is half of the chain.
    The system is initialized in (a) a charge-density-wave %(CDW) 
    state, $|\psi_0\rangle = (\prod_{j=1}^{N/2} \hat{f}_{2j}^\dagger) |0\rangle$;
    (b) %The system is initialized in 
    a randomly sampled (over 10000 realizations) mixed Gaussian state. %The maker shows the average over 10000 samples.
    }
    \label{fig:dynamical_area_law}
\end{figure}

%Although there are 
We emphasize a fundamental difference from similar statements for entanglement entropy in pure states in Refs.~\cite{Bravyi-Lieb-RobinsonBoundsOrder-2006h,Gong-EntanglementAreaSystems-2017e} based on the SIE theorems. Therein, 
the area law of entanglement entropy rate holds for both \textit{increase} and \textit{decrease} rates, as the SIE theorem applies to both. %for both of these. 
In contrast, here the area law of negativity change rate should only be true for the \textit{increase} rate, as shown in Fig.~\ref{fig:dynamical_area_law}. 
This difference manifests in the difference between the bound for the increase rate \eqref{negativity_increase_rate_bound} and that %the bound 
for the magnitude %of change rate 
\eqref{negativity_rate_bound}: while the former grows proportionally to the boundary size %perimeter 
as far as the system is sufficiently local, the latter becomes proportional to the system size. 
For example, if one prepares a volume-law entangled initial state and considers %on-site decoherence or
particle-loss dynamics, %at every site, 
then negativity clearly decreases proportionally to the subsystem size, not its boundary. %However, 
Nevertheless, regarding the \textit{increase} rate, we can prove an area law regardless of the initial state.

To prove the area law \eqref{negativity_rate_area_law_statement}, we set the coefficients of Lindblad operators as $L_{\mu,j} = L_{(\bm{r},\nu),(\boldsymbol{r}',s)%j
}$, with $\nu\in J$ labeling the local dissipation channels. Let us impose the locality of dissipation as 
\begin{align}
    \|\Pi_{\boldsymbol{r}}L\Pi_{\boldsymbol{r}'}\|\le lF_\alpha(|\boldsymbol{r}-\boldsymbol{r}'|)\label{dissipation_locality}
\end{align}
in addition to the locality of Hamiltonian \eqref{locality}, where $l$ is an $\mathcal{O}(1)$ constant and $L$ denotes the matrix of coefficients $L_{\mu,j}$.
Assuming $\alpha>D$, this assumption %of locality 
\eqref{dissipation_locality} ensures the locality of $X,Y$ as
\begin{align}
    \|\Pi_{\bm{r}}X\Pi_{\bm{r}'}\|,\|\Pi_{\bm{r}}Y\Pi_{\bm{r}'}\| \leq 2c'lF_\alpha(|\bm{r}-\bm{r}'|)\label{dynamics_locality}
\end{align}
%with an $\mathcal{O}(1)$ constant $c'$, 
by applying the self-production property $\sum_{\bm{r}} F_\alpha(|\bm{r}_1-\bm{r}|)F_\alpha(|\bm{r}-\bm{r}_2|)\leq c'F_\alpha(|\bm{r}_1-\bm{r}_2|)$ \cite{Nachtergaele-PropagationCorrelationsSystems-2006h}, where $c'$ is an $\mathcal{O}(1)$ constant. In this case, the area law holds if $\sum_{\bm{r}\in A, \bm{r}' \in B} F_\alpha(|\bm{r}-\bm{r}'|)$ is convergent. Since the summation can be evaluated in the same way as Eq.~\eqref{area_law_eq_state}, $F_\alpha(r)$ should decay faster than $r^{-(D+1)}$ (i.e., $\alpha>D+1$). %We 
Note that the %obtained 
same condition was obtained %is the same as 
for the dynamical area law of entanglement entropy shown in Ref.~\cite{Gong-EntanglementAreaSystems-2017e}. %As in the case of unitary dynamics, it remains open to determine the optimal condition for this dynamical area law in long-range systems.

Our dynamical area law adds new insight into dissipation engineering. While engineered %to the idea of utilizing dissipation as a resource. 
%In modern physics, 
dissipation has been %is 
widely recognized as a quantum resource %tool rather than an obstacle for quantum information processing 
\cite{Poyatos-QuantumReservoirIons-1996v,Harrington-EngineeredDissipationScience-2022z,Verstraete-QuantumComputationDissipation-2009b}, %. However, 
it remains to understand whether %has been indicated that 
introducing dissipation could surpass the unitary limit  %does not improve the efficiency of preparing long-range entangled states 
\cite{Konig-GeneratingTopologicalDissipation-2014o}. Our dynamical area law gives a no-go result regarding generation of bipartite entanglement. The absence of nonunitary advantage %explains this result from a fundamental law of physics which governs the nonunitary entanglement dynamics. 
%We indicate that this consequence 
is already suggested %in 
by the similarity between the SIE and the bound for the increase change rate of entanglement \eqref{negativity_increase_rate_bound}. %: the coincidence of this bound with the SIE theorem somehow suggests that the fundamental constraint of entanglement generation cannot be removed even in nonunitary dynamics, as far as the system exhibits locality.

\textit{Summary and outlook.---} Focusing on free fermions, we have proposed the twisted characteristic polynomial to
directly relate the entanglement negativity to the covariance matrix. Our formalism allows us to derive universal upper and lower bounds on negativity, which unveil %are tight for high-temperature thermal states. An important implication of these bounds is 
the connection between the clustering of correlations %in free-fermion states 
and the entanglement area law. 
Moreover, we establish the first dynamical area law governing the %rate of 
entanglement negativity growth in local dissipative dynamics, revealing a fundamental universal constraint on entanglement generation beyond the unitary paradigm.

Several %promising 
directions emerge for future work. One direction is to apply our formalism to %symmetry-protected topological phases in mixed states.
topological insulators/superconductors at finite temperatures. In particular,  %Since negativity has been found to detect quantum phase transitions~\cite{Shapourian-PartialTime-reversalSystems-2017q,Choi-Finite-temperatureEntanglementDimension-2024f,Wen-TopologicalEntanglementTheories-2016h,Castelnovo-NegativityTopologicalCode-2013s,Lu-DetectingTopologicalNegativity-2020q}, 
it is %an 
intriguing %direction 
to study how zeros of the twisted characteristic polynomial, as the counterpart of the entanglement spectrum~\cite{Fidkowski-EntanglementSpectrumSuperconductors-2010r}, are related to topological phases. %representing the negativity  
%Our analytical formula will be useful for investigating the properties of this "topological negativity" which should remain finite at finite temperature, although our bounds are not optimal enough in this regime. 
Furthermore, our framework can be applied to the entanglement structure of topological states that appear as steady states of engineered dissipative dynamics~\cite{Bardyn-TopologyDissipation-2013j}. %Despite the pure state entanglement spectrum characterizes the ground state topological order~\cite{Fidkowski-EntanglementSpectrumSuperconductors-2010r}, its nonequilibrium counterpart remains unclear. 
%Therefore, 
In particular, it is interesting to investigate how negativity characterizes these nonequilibrium topological phases and how it evolves under such dynamics. 

Another natural question is whether our results can be generalized to interacting fermions. %In the Gibbs state case, 
Very recently, it has been proved that general Gibbs states are %probabilistic 
mixtures of Gaussian states at %above sufficiently 
high temperatures~\cite{Akshar-High-TemperatureFermionicStates-2025s}. Therefore, we expect that %derive bounds for the Gibbs states of 
the rigorous analysis of general interacting fermions may still benefit from %fermionic Hamiltonians based on 
our bounds. 
%Moreover, our dynamical area law would be extended to interacting systems based on the Lieb-Robinson bound in open quantum systems

Finally, a profound question concerns the meaning of ``twist''. Since %the structure of 
our twisted characteristic polynomial can be obtained from %viewed as 
a partial modular transformation (i.e., $\lambda\to-\lambda^{-1}$) of the ordinary characteristic polynomial, it would be %an 
interesting %challenge 
to generalize our formula to $\mathbb{Z}_n$ parafermionic systems \cite{Fendley-FreeParafermions-2014j, Alicea-TopologicalPhasesBlueprints-2016p} and to multipartite settings, where more blocks are expected to be twisted in different ways.

%%%=========== main text end ===========%%%

%%% acknowledgments %%%
\begin{acknowledgments}
  R.~M. thanks Masahiro Hoshino for valuable discussions, especially those that helped clarify the arguments regarding entanglement dynamics, and helpful comments on the manuscript. Z.G. acknowledges support from the University of Tokyo Excellent Young Researcher Program and from JST ERATO Grant No. JPMJER2302, Japan.
\end{acknowledgments}

%%% bibliography %%%
\bibliography{paperpile}

%%%=========== Appendix ===========%%%
\onecolumngrid
\clearpage

\begin{center}
  \textbf{\large Supplemental Material for ``Entanglement negativity in free fermions:\\twisted characteristic polynomial, universal bounds and area laws''}
\end{center}

\beginsupplement

\section{Twisted characteristic polynomial}\label{appendix_negativity_formula}

We review the logarithmic negativity for fermions proposed in~\cite{Shapourian-PartialTime-reversalSystems-2017q}, and show the derivation of our formula from the covariance matrix corresponding to the twisted partial transpose introduced in Ref.~\cite{Shapourian-TwistedUntwistedFermions-2019u}.

\subsection{Untwisted and twisted partial transpose}

To state the definition of fermionic partial transpose, we begin by considering a general fermionic system with $N$ modes. The Hilbert space $\mathcal{H}$ is spanned by the Fock basis $\ket{n_1,n_2\cdots, n_N}$, where $n_j=0,1$ is the occupation number of the $j$th mode. Any linear operator can be expressed in terms of creation and annihilation operators $\hat{f}_j^\dagger$ and $\hat{f}_j$, 
which act on the Fock basis as
\begin{align}
    &\hat{f}_j \ket{n_1, \cdots, n_j,\cdots n_N} \notag\\
    &\quad =\delta_{n_j,1} (-1)^{\sum_{i=1}^{j-1}n_i} \ket{n_1, \cdots, n_{j-1},0,n_{j+1},\cdots n_N},\notag\\
    &\hat{f}_j^\dagger \ket{n_1, \cdots, n_j,\cdots n_N} \notag\\
    &\quad =\delta_{n_j,0} (-1)^{\sum_{i=1}^{j-1}n_i} \ket{n_1, \cdots, n_{j-1},1,n_{j+1},\cdots n_N}. \label{fermioncom}
\end{align} 
Due to \eqref{fermioncom}, $\hat{f}_j^\dagger$ and $\hat{f}_j$ satisfy the anticommutation relations
\begin{equation}
    \{\hat{f}_i,\hat{f}_j\} = \{\hat{f}_i^\dagger, \hat{f}_j^\dagger\}=0\;, \{\hat{f}_i,\hat{f}_j^\dagger\}=\delta_{ij},
\end{equation}
where $\{\hat{A},\hat{B}\}=\hat{A}\hat{B}+\hat{B}\hat{A}$.

For the definition of fermionic partial transpose, it is convenient to introduce a Majorana representation. Majorana operators $\hat{c}_1,\cdots,\hat{c}_{2N}$ are constructed from the fermionic creation and annihilation operators as
\begin{equation}
    \hat{c}_{2j-1} = \hat{f}_j +\hat{f}_j^\dagger\;,\;\hat{c}_{2j} = -i(\hat{f}_j-\hat{f}_j^\dagger),
\end{equation}
and satisfy $\{\hat{c}_i,\hat{c}_j\} = 2\delta_{ij}$. Any physical operator $\hat{X}$ acting on $\mathcal{H}$ is expressed in terms of polynomials of $\hat{c}_j$'s,
\begin{equation}
    X =\sum_{k=1}^{2N} \sum_{p_1<p_2<\cdots<p_k,k=\text{even}} X_{p_1\cdots p_k} \hat{c}_{p_1} \cdots \hat{c}_{p_k},
\end{equation}
where the restriction $k=\text{even}$ comes from the superselection rule.

To study the entanglement in a bipartite system, we divide the system into two subsystems A and B. The total Hilbert space is $\mathcal{H} = \mathcal{H}_A \otimes \mathcal{H}_B$, where $\mathcal{H}_A$ is generated by $\hat{c}_j$ with $j=1,\cdots,2N_A$ in subsystem A and $\mathcal{H}_B$ by $j=2N_A+1,\cdots, 2N(=2(N_A+N_B))$ in subsystem B. Any physical state on $\mathcal{H}$ is given by 
\begin{equation}
    \hat{\rho} = \sum_{k_1,k_2}^{k_1+k_2=\text{even}} \rho_{p_1\cdots p_{k_1},q_1\cdots q_{k_2}} \hat{a}_{p_1} \cdots \hat{a}_{p_{k_1}} \hat{b}_{q_1} \cdots \hat{b}_{q_{k_2}},
\end{equation}
where $\{\hat{a}_j\}$ and $\{\hat{b}_j\}$ are majorana operators acting on $\mathcal{H}_A$ and $\mathcal{H}_B$, respectively.
Then, the \textit{untwisted} partial transpose is defined as 
\begin{equation}
    \hat{\rho}^{\rm T_A} = \sum_{k_1,k_2}^{k_1+k_2=\text{even}} \rho_{p_1\cdots p_{k_1},q_1\cdots q_{k_2}} i^{k_1}\hat{a}_{p_1} \cdots \hat{a}_{p_{k_1}} \hat{b}_{q_1} \cdots \hat{b}_{q_{k_2}}.
\end{equation}
This \textit{untwisted} partial transpose preserves the tensor product structure (and thus the corresponding logarithmic negativity satisfies additivity) but not Hermiticity,
\begin{equation}
    \qty(\hat{\rho}^{\rm T_A})^\dagger = (-1)^{\hat{F}_A} \hat{\rho}^{\rm T_A} (-1)^{\hat{F}_A},
\end{equation}
where $\hat{F}_A = \sum_{j \in A} \hat{f}^\dagger_j \hat{f}_j$ is the number operator in A. This motivates us to introduce \textit{twisted} partial transpose~\cite{Shapourian-TwistedUntwistedFermions-2019u}
\begin{equation}
   \hat{\rho}^{\tilde{\rm T}_A} = \hat{\rho}^{\rm T_A} (-1)^{\hat{F}_A},
\end{equation}
which is Hermitian by definition. Since $\|\hat{\rho}^{\rm T_A}\|_1=\|\hat{\rho}^{\tilde{\rm T}_A}\|_1$, it suffices to know the spectrum of $\hat{\rho}^{\tilde{\rm T}_A}$ to calculate logarithmic negativity.

\subsection{Covariance matrix of the twisted partial transpose}

In the Gaussian case, it is rather convenient to apply twist partial transpose for the corresponding covariance matrix, and calculate logarithmic negativity from the covariance matrix for the twisted partial transpose $\tilde{\Gamma}$. Indeed, they are related as~\cite{Shapourian-TwistedUntwistedFermions-2019u}
\begin{align}
    \mathcal{E} &= \ln |\Pf(\Gamma_A)| + \sum_{j=1}^N \ln \max\{1,\tilde{\lambda}_j\},\label{negativity_formula_scipost}
\end{align}
where $\pm \tilde{\lambda}_j\in\mathbb{R}\;(j=1,2,\cdots,N)$ are eigenvalues of $\tilde{\Gamma}$. Here we distinguish $\tilde{\lambda}_j$ from ${\lambda}_j$, which are the roots of the twisted characteristic polynomial $P(\lambda)$, but later it turns out that they are the same.

Let us move to calculate the explicit form of the covariance matrix for the \textit{twisted} partial transpose $\tilde{\Gamma}$, based on the \textit{untwisted} version given in Ref.~\cite{Shapourian-TwistedUntwistedFermions-2019u}. We use the same bipartition for the original covariance matrix as \eqref{cov_partition} in the main text:
\begin{align}
    \Gamma = \begin{pmatrix}
        \Gamma_{A} & \Gamma_{AB}\\
        \Gamma_{BA} & \Gamma_{B}
    \end{pmatrix}.
\end{align}
Then, the covariance matrix of untwisted partial transpose $\hat{\rho}^{\rm T_A}$ and $\hat{\rho}{^{\rm T_A}}^\dagger$ is given by~\cite{Shapourian-TwistedUntwistedFermions-2019u}
\begin{align}
    \Gamma_{\pm} = \begin{pmatrix}
        -\Gamma_A & \pm i\Gamma_{AB}\\
        \pm i\Gamma_{BA} & \Gamma_B
    \end{pmatrix},
\end{align}
where $\pm$ correspond to $\hat{\rho}^{\rm T_A}$ and $\hat{\rho}{^{\rm T_A}}^\dagger$. As for the twisted partial transpose $\hat{\rho}^{\tilde{\rm T}_A} = \hat{\rho}^{\rm T_A} (-1)^{\hat{F}_A}$ and $\hat{\rho}^{{\rm T_A}^\dagger} (-1)^{\hat{F}_A} $, we have~\cite{Shapourian-TwistedUntwistedFermions-2019u}
\begin{align}
    e^{\tilde{\Omega}_\pm} = \frac{\matidentity + \Gamma_{\pm}}{\matidentity -\Gamma_{\pm}}U_A,\;\;\;U_A = \begin{pmatrix}
        -\matidentity_A & 0\\
        0 & \matidentity_B
    \end{pmatrix}.\label{twistedgaussianmat}
\end{align}
Substituting Eq.~\eqref{twistedgaussianmat} into $\tilde{\Gamma}_{\pm} = \tanh \frac{\tilde{\Omega}_{\pm}}{2}$, we obtain 
\begin{align}
    \tilde{\Gamma} = \frac{e^{\tilde{\Omega}_\pm}-\mathbb{I}}{e^{\tilde{\Omega}_\pm}+\mathbb{I}} &= [(\mathbb{I}-\Gamma_\pm)^{-1}(\mathbb{I}+\Gamma_\pm)U_A+\mathbb{I}]^{-1} [(\mathbb{I}-\Gamma_\pm)^{-1}(\mathbb{I}+\Gamma_\pm)U_A-\mathbb{I}]\notag\\
    &= [(\mathbb{I}+\Gamma_\pm)U_A+(\mathbb{I}-\Gamma_\pm)]^{-1}[(\mathbb{I}+\Gamma_\pm)U_A-(\mathbb{I}-\Gamma_\pm)]\notag\\
    &= (\Pi_B-\Gamma_\pm \Pi_A)^{-1} (\Gamma_\pm \Pi_B-\Pi_A),\label{A1}
\end{align}
where we used $2\Pi_A=\mathbb{I}-U_A$ and $2\Pi_B = \mathbb{I} + U_A$. Since $\Pi_{A,B}$ is the projector onto $A, B$, we can explicitly write down the block forms as
\begin{align}
    \Pi_B-\Gamma_\pm \Pi_A = \begin{pmatrix}
        \Gamma_A & 0\\
        \mp i\Gamma_{BA} & \mathbb{I}_B
    \end{pmatrix},
    \Gamma_\pm \Pi_B - \Pi_A= \begin{pmatrix}
        -\mathbb{I}_A & \pm i\Gamma_{AB}\\
        0 & \Gamma_B
    \end{pmatrix}.\label{A2}
\end{align}
Assuming the invertibility of $\Gamma_A$, the inverse of the former turns out to be 
\begin{align}
    (\Pi_B-\Gamma_\pm \Pi_A)^{-1} = \begin{pmatrix}
        \Gamma_A^{-1} & 0\\
        \pm i \Gamma_{BA} \Gamma_A^{-1} &\mathbb{I}_B
    \end{pmatrix}.\label{A3}
\end{align}
Therefore, by substituting Eqs.~\eqref{A2} and \eqref{A3} into Eq.~\eqref{A1}, we can derive the covariance matrix corresponding to $\hat{\rho}^{\tilde{\rm T}_A} = \hat{\rho}^{\rm T_A} (-1)^{\hat{F}_A}$ and $\hat{\rho}^{{\rm T_A}^\dagger} (-1)^{\hat{F}_A} $ as
\begin{align}
    \tilde{\Gamma}_{\pm} = \begin{pmatrix}
        -\Gamma_{A}^{-1} & \pm i\Gamma_{A}^{-1} \Gamma_{AB}\\
        \mp i\Gamma_{BA} \Gamma_{A}^{-1} & \Gamma_B -\Gamma_{BA} \Gamma_A^{-1} \Gamma_{AB}
    \end{pmatrix}.
\end{align}

\subsection{Derivation of twisted characteristic polynomial}

Finally, we show the equivalence of the formula \eqref{negativity_formula_scipost} and our formula \eqref{negativity_formula}, under the assumption of the invertibility of $\Gamma_A$. Using the invariance of determinant under column linear operations (or right multiplying the upper triangle matrix $\begin{pmatrix}
    \matidentity_A & \mp i\Gamma_{AB}\\
    0 & \matidentity_B
\end{pmatrix}$ with unit determinant), we have
\begin{align}
    \det (\tilde{\Gamma}_{\pm}-\lambda\matidentity) = \det \begin{pmatrix}
        -\Gamma_A^{-1}-\lambda\matidentity_A & \mp i\lambda \Gamma_{AB}\\
        \mp i\Gamma_{BA}\Gamma_A^{-1} & \Gamma_B-\lambda \matidentity_B
    \end{pmatrix}.
\end{align}
Note that
\begin{align}
    \begin{pmatrix}
        -\Gamma_A^{-1}-\lambda\matidentity_A & \mp i\lambda \Gamma_{AB}\\
        \mp i\Gamma_{BA}\Gamma_A^{-1} & \Gamma_B-\lambda \matidentity_B
    \end{pmatrix} = \begin{pmatrix}
        \lambda \matidentity_A &0\\
        0 &\pm i\matidentity_B
    \end{pmatrix} \begin{pmatrix}
        \Gamma_A +\lambda^{-1}\matidentity_A & \Gamma_{AB}\\
        \Gamma_{BA} & \Gamma_B-\lambda \matidentity_B
    \end{pmatrix} \begin{pmatrix}
        -\Gamma_A^{-1} & 0\\
        0 & \mp i\matidentity_B
    \end{pmatrix},
\end{align}
as long as $\Gamma_A$ is invertible. This implies
\begin{align}
    P(\lambda) = \det(\Gamma_A)\det (\tilde{\Gamma}_{\pm}-\lambda\matidentity),
\end{align}
where $P(\lambda)$ is the twisted characteristic polynomial given in Eq.~\eqref{tcp}. One can see that the coefficient of the leading term $\lambda^{2N}$ is exactly $\det (\Gamma_A) = \Pf(\Gamma_A)^2$, according to which Eq.~\eqref{negativity_formula} follows. As the coefficients in $P(\lambda)$ depend continuously on $\Gamma_A$, the zeros should also have a continuous dependence. This justifies the continuity of Eq.~\eqref{negativity_formula} in terms of $\Gamma_A$ regardless of its invertibility.

\subsection{two modes example}

We demonstrate our formula in the simplest two fermionic modes example. First we treat Gibbs state given by the Hamiltonian
\begin{align}
    \hat{H} = -J(\hat{f}_2^\dagger \hat{f}_1 + \HC), \label{two_mode_hamiltonian}
\end{align}
which is also treated in appendix of Ref.~\cite{Shapourian-Finite-temperatureEntanglementFermions-2019y}. One can check that the Gibbs state is determined by $\Omega = \beta J\sigma_x\otimes\sigma_y$, so that $\Gamma = \tanh(\frac{\beta J}{2}) \sigma_x\otimes\sigma_y$. The twisted characteristic polynomial for the two-mode entanglement reads
\begin{align}
    P(\lambda) = \qty(1+\tanh^2\qty(\frac{\beta J}{2}))^2 \lambda^2,
\end{align}
whose roots (simply 0) are smaller than 1. This implies
\begin{align}
    \mathcal{E}(\beta) = \ln\qty(\frac{2\cosh(\beta J)}{1+\cosh(\beta J)}).\label{two_mode_Gibbs_negativity}
\end{align}
This result is exactly the same in Ref.~\cite{Shapourian-Finite-temperatureEntanglementFermions-2019y}, where this result is derived by expanding $\hat{{\rho}}$ in fock basis and directly apply the partial transpose in each matrix elements. The asymptotic form in the high temperature limit reads $\mathcal{E}(\beta) = (\beta J)^2/4 + \mathcal{O}\qty((\beta J)^4)$, which is consistent with the universal scaling obtained by our bounds.

Next, we treat the dissipative dynamics of two fermionic modes, which can be thought as a counterpart of \eqref{two_mode_hamiltonian} in the sense of entanglement sudden death. We take the initial entangled state $\frac{1}{\sqrt{2}}(\hat{f}_1^\dagger+\hat{f}_2^\dagger)\ket{0}$, whose covariance matrix is 
\begin{align}
    \Gamma_0 = \sigma_x \otimes \sigma_y = \begin{pmatrix}
        0 & \sigma_y \\
        \sigma_y & 0
    \end{pmatrix}.
\end{align}
Note that this initial state is nothing but the ground state of the previous two-mode Hamiltonian \eqref{two_mode_hamiltonian}.
The dynamics is described by the Lindblad operators
$\hat{L}_1=\sqrt{\gamma_1}\hat{f}_1$ and $\hat{L}_2=\sqrt{\gamma_2}\hat{f}_2$ with decay rates $\gamma_1$ and $\gamma_2$, and we take $\gamma_1=\gamma_2=\gamma$ to simplify our calculation. Then the $X,Y$ defined in Eq.~\eqref{cov_dynamics} is $X = \frac{\gamma}{2}\matidentity_4, Y = -\frac{\gamma}{2}\sigma_y^{\oplus 2}$. Therefore
\begin{align}
    \Gamma(t) = e^{-\gamma t}(\Gamma_0+\sigma_y^{\oplus 2}) -\sigma_y^{\oplus 2} = \begin{pmatrix}
        -\sigma_y(1-e^{-\gamma t}) & \sigma_y e^{-\gamma t}\\
        \sigma_y e^{-\gamma t} & -\sigma_y(1-e^{-\gamma t})
    \end{pmatrix},
\end{align}
which yields the twisted characteristic polynomial
\begin{align}
    P(\lambda;t) = -(1-e^{-\gamma t})^2 \qty[\lambda^4 - 2 \qty(\frac{2e^{-2\gamma t}}{(1-e^{-\gamma t})^2}+1)\lambda^2+1]. \label{two_mode_dynamics_poly}
\end{align}
In $t>0$, the roots of \eqref{two_mode_dynamics_poly} are 
\begin{align}
    \lambda_\pm(t) = \sqrt{b(t) \pm \sqrt{b(t)^2-1}}, \;b(t) = \frac{2e^{-2\gamma t}}{(1-e^{-\gamma t})^2}+1,
\end{align}
and $0 < \lambda_-(t) < 1 < \lambda_+(t)$ because $b(t)>1$. Thus, the exact solution of negativity is 
\begin{align}
    \mathcal{E}(t) &= \frac{1}{2}\ln |P_>(0;t)| = \frac{1}{2} \ln \qty[(1-e^{-t})^2(b(t) + \sqrt{b(t)^2-1})]\notag\\
    &= \frac{1}{2}\ln \qty[2e^{-2\gamma t} + (1-e^{-\gamma t})^2 + 2e^{-\gamma t} \sqrt{e^{-2\gamma t} + (1-e^{-\gamma t})^2}].\label{two_mode_exact_dynamics}
\end{align}
Since \eqref{two_mode_exact_dynamics} satisfies $\mathcal{E}(0)=\ln2$, this is true for $t\geq 0$. We can derive the asymptotic form of \eqref{two_mode_exact_dynamics} as
\begin{align}
    \mathcal{E}(t) = 
    \begin{cases}
        \ln 2 - \gamma t + \mathcal{O}((\gamma t)^2) & \text{when $t \to 0$,}\\
        \frac{1}{2}e^{-2\gamma t} + \mathcal{O}( e^{-3\gamma t})& \text{when $t \to \infty$.}
    \end{cases}
\end{align}

\begin{figure}[ht]
	\centering
	\includegraphics[width=0.5\textwidth]{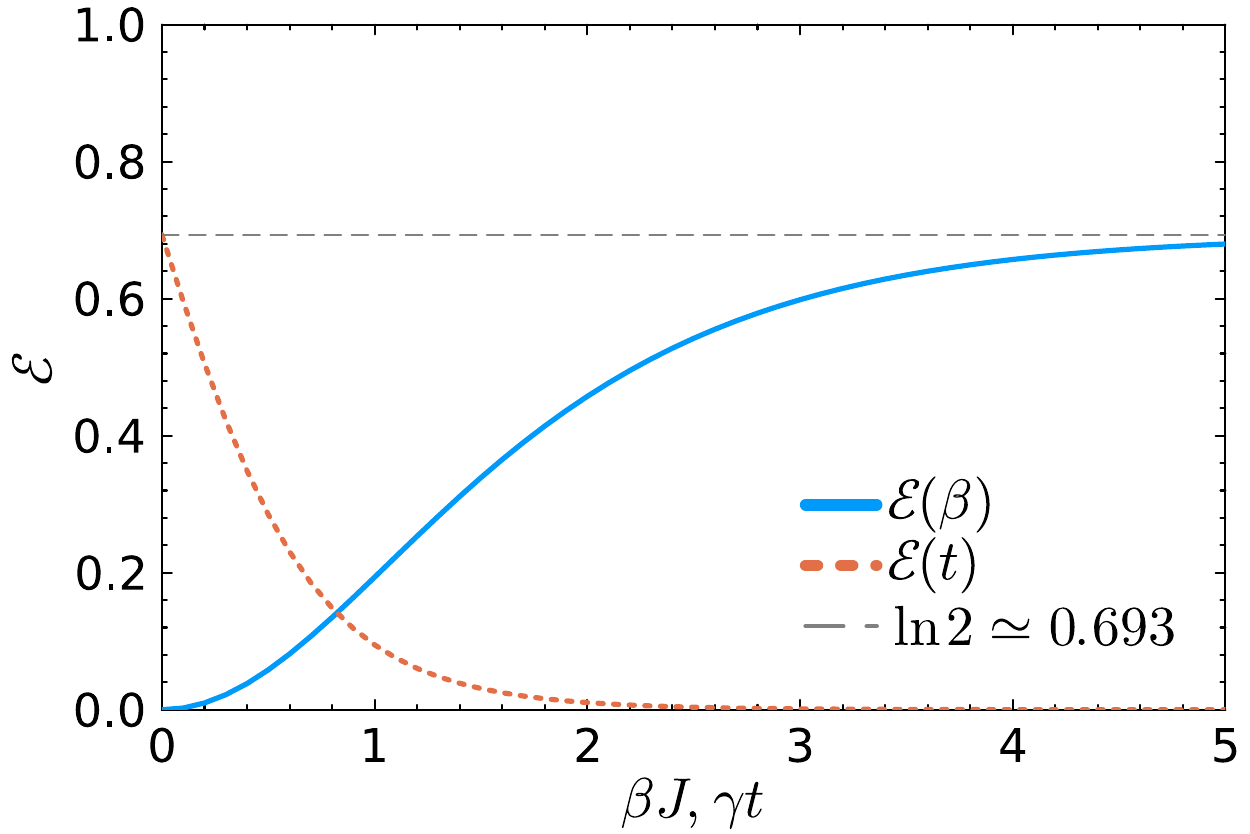}
	\caption{Exact negativity of the Gibbs state \eqref{two_mode_Gibbs_negativity} and of the dissipative dynamics \eqref{two_mode_exact_dynamics}.}
	\label{fig:two_modes_example}
\end{figure}

We show these two cases in Fig.~\ref{fig:two_modes_example}. %This illustration manifests the properties we discussed in the main text by using the universal bounds.

\section{Universal bounds on negativity}

\subsection{Derivation of the bounds}

In this section, we show how to derive the upper and lower bounds of negativity, Eq.~\eqref{upperbound} and \eqref{lowerbound} in the main text.

A general Gaussian operation is in the form of Eq.~\eqref{gaussian_operation}, and here we consider the specific Gaussian channel with $D=0$ and $K=i\sqrt{k}\matidentity \;(k\geq0)$ such that
\begin{align}
    \Gamma_{\text{out}} = k\Gamma_{\text{in}} + C.
\end{align}
The constraint in Eq.~\eqref{CP_constraint} reads
\begin{align}
    \left|\left|\frac{\sigma_z+\sigma_0}{2}\otimes C - \sqrt{k}\sigma_y\otimes\matidentity\right|\right| = \frac{1}{2} \qty(\|C\|+\sqrt{\|C\|^2+4k}) \leq 1,
\end{align}
which is equivalent to
\begin{align}
    k \leq 1- \|C\|, \|C\| \leq 1.
\end{align}
To derive the upper bound \eqref{upperbound}, we choose
\begin{align}
    \Gamma_{\text{in}} = k^{-1} \begin{pmatrix}
        0 & \Gamma_{AB}\\
        \Gamma_{BA} & 0
    \end{pmatrix}, C = \Gamma_A \oplus \Gamma_B,
\end{align}
so that $\Gamma_{\text{out}}$ is exactly the state of interest. In this case, $\|C\| = \max\{\|\Gamma_A\|, \|\Gamma_B\|\}$ and there is additional constraint $\|\Gamma_{\text{in}}\|\leq1 \Leftrightarrow \|\Gamma_{AB}\|\leq k$. The logarithmic entanglement negativity of the input state is given by
\begin{align}
    \mathcal{E}_{\text{in}} = \frac{1}{2}\tr[\ln(\matidentity_B+k_-^{-2}\Gamma_{BA}\Gamma_{AB})],
\end{align}
so the tightest version is $k=1-\max\{\|\Gamma_A\|, \|\Gamma_B\|\}$ with further constraint $k_- \leq \|\Gamma_{AB}\|$.

To derive the lower bound \eqref{lowerbound}, we choose
\begin{align}
    \Gamma_{\text{in}} = \Gamma, C=-k(\Gamma_A\oplus \Gamma_B).
\end{align}
In this case, $\|C\| = k\max\{\|\Gamma_A\|, \|\Gamma_B\|\}$ and the only constraint is
\begin{align}
    k \leq (1+\max{\|\Gamma_A\|, \|\Gamma_B\|})^{-1}.
\end{align}
The lower bound is given by the entanglement negativity of the output state:
\begin{align}
    \mathcal{E}_{\text{out}} = \frac{1}{2}\tr[\ln(\matidentity_B+k_+^{2}\Gamma_{BA}\Gamma_{AB})]
\end{align}
which is maximized by $k=k_+^{-1}$.

We emphasize these bounds should be far from optimal because our choices are rather specific (yet make the calculations extremely simple). For example, there is an obvious improvement for the lower bound by choosing C to be either $-k \Gamma_A \oplus 0$ or
$0 \oplus (-k\Gamma_B)$, so that $k_+$ in Eq.~\eqref{lowerbound} can be replaced by $1 + \min\{\|\Gamma_A\|, \|\Gamma_B\|\}$.

\subsection{Numerical calculation of the bounds}

For the numerical verification of our bounds, we employed two different models and calculated the bounds and exact negativity of the Gibbs state of those models. The models are tight-binding model and Kitaev chain with open boundary condition, whose Hamiltonians are given by
\begin{align}
    \hat{H}_\text{TB} &= -\sum_{i=1}^{N-1} t (\hat{f}^\dagger_i\hat{f}_{i+1} + \HC)=\sum_{i=1}^{N-1}\frac{it}{2} (\hat{c}_{2i}\hat{c}_{2i+1}-\hat{c}_{2i-1}\hat{c}_{2i+2})\label{TB},\\ \hat{H}_\text{Kitaev} &=  -\sum_{i=1}^{N-1} t (\hat{f}^\dagger_i - \hat{f}_i)(\hat{f}_{i+1}^\dagger+\hat{f}_{i+1}) - 2\sum_{i=1}^N \qty(\hat{f}_i^\dagger \hat{f}_i - \frac{1}{2})= -\sum_{i=1}^N i\hat{c}_{2i-1}\hat{c}_{2i} + \sum_{i=1}^{N-1} it \hat{c}_{2i}\hat{c}_{2i+1}.\label{Kitaev}
\end{align}
In the main text, we showed that our bound exactly captures the universal behavior of the free-fermion negativity in the high temperature regime. However, we note that our bounds \eqref{simplerbounds} are far from optimal in the low temperature regime. %, as in Fig.~\ref{fig:bound_low_temp}. 
We observed that the behavior is qualitatively the same if we employ tighter bounds \eqref{upperbound} and \eqref{lowerbound}. We expect that bounds can be further improved by making a better choice of Gaussianity-preserving local operations.

\section{The change rate of negativity}

\subsection{Derivation of the formula for the change rate of negativity}

We show the derivation of the explicit formula of negativity change rate $\partial_t \mathcal{E}(t)$. We write time dependence of the twisted characteristic polynomial \eqref{tcp} as $P(\lambda;t) = \det(\Gamma_A(t)) \prod_{j=1}^{2N}(\lambda-\lambda_j(t))$. Then, the time derivative of Eq.~\eqref{negativity_formula} reads
\begin{align}
    \pdv{\mathcal{E}}{t} &= \frac{1}{2}\qty(\frac{\partial_t\det(\Gamma_A(t))}{\det(\Gamma_A(t))} + \sum_{j:|\lambda_j|>1}\frac{\partial_t \lambda_j(t)}{\lambda_j(t)})\notag\\
    &= \frac{1}{2}\int_{-\pi}^{\pi} \frac{dk}{2\pi} \frac{\partial_t P(e^{-ik};t)}{P(e^{-ik};t)} = \frac{1}{2}\int_{-\pi}^{\pi} \frac{dk}{2\pi} \partial_t \ln P(e^{-ik};t),
\end{align}
where we used an integral
\begin{align}
    \int_{-\pi}^{\pi} \frac{dk}{2\pi} \frac{1}{\lambda_j-e^{-ik}} = \begin{cases}
    0 & \text{if $|\lambda_j|<1$,} \\
    \frac{1}{\lambda_j} & \text{if $|\lambda_j|>1$}.
  \end{cases}
\end{align}
To proceed, we define matrix $\Xi$ by $P(\lambda;t) = \det \Xi(\lambda;t)$. Since
\begin{align}
    \partial_t \ln P(e^{-ik};t) &= \partial_t \ln\det \Xi(e^{-ik};t)\notag \\
    &= \Tr[\Xi^{-1}(e^{-ik};t) \pdv{t} \Xi(e^{-ik};t)],
\end{align}
and $\Xi$ is related to the covariance matrix $\Gamma$ via
\begin{align}
    \Xi(\lambda;t)&=U(\lambda)(V(\lambda)+\Gamma(t)),\notag\\
    U(\lambda)&=\begin{pmatrix}
        \lambda \matidentity_A & 0\\
        0 & \matidentity_B
    \end{pmatrix}, V(\lambda) = \begin{pmatrix}
        \lambda^{-1} \matidentity_A & 0\\
        0 & -\lambda \matidentity_B\\
    \end{pmatrix},
\end{align}
we arrive at the explicit formula of negativity rate:
\begin{align}
    \pdv{\mathcal{E}}{t} =  \frac{1}{2}\int_{-\pi}^{\pi} \frac{dk}{2\pi} \Tr[\qty(V(e^{-ik}) + \Gamma(t))^{-1} \pdv{\Gamma(t)}{t}].
\end{align}
Therefore, denoting $Z_{AB}(k) = V(e^{-ik}) =e^{ik}\matidentity_A \oplus(-e^{-ik})\matidentity_B$, we derive Eq.~\eqref{negativity_rate_formula} in the main text:
\begin{equation}
\pdv{\mathcal{E}}{t} =\frac{1}{2}\Tr\left[\mathcal{P}_{AB}(\Gamma) \pdv{\Gamma}{t} \right],\;\;\; \mathcal{P}_{AB} (\Gamma) = \int^\pi_{-\pi} \frac{dk}{2\pi} (Z_{AB}(k) + \Gamma)^{-1}.
\label{dEdt}
\end{equation}

\subsection{Block diagonal representation}

To gain further insights into $\mathcal{P}_{AB}$, it turns out to be helpful to go back to the formalism presented in Ref.~\cite{Shapourian-TwistedUntwistedFermions-2019u}. We first rewrite $\mathcal{P}_{AB}(\Gamma)$ into a contour integral along a unit circle:
\begin{equation}
\mathcal{P}_{AB}(\Gamma) = \oint_{|z|=1} \frac{dz}{2\pi i} 
\begin{bmatrix}
z^2+z\Gamma_A & z\Gamma_{AB} \\
z\Gamma_{BA} & z\Gamma_B -1
\end{bmatrix}^{-1}.
\end{equation}
By slightly deforming the integral contour, we can make the integrand matrix invertible and apply the matrix inverse formula for block matrices. The diagonal block restricted to $A$ reads
\begin{equation}
\begin{split}
P_A \mathcal{P}_{AB}(\Gamma) P_A &= \oint_{|z|=1} \frac{dz}{2\pi i} (z^2+z\Gamma_A- z^2\Gamma_{AB}(z\Gamma_B - 1)^{-1}\Gamma_{BA})^{-1} \\
&=\oint_{|\zeta|=1}\frac{d\zeta}{2\pi i} (1+\Gamma_{AB}\Gamma_{BA}+\zeta\Gamma_A - \Gamma_{AB}(1-\zeta\Gamma_B^{-1})^{-1}\Gamma_{BA})^{-1} \\
&=\oint_{|\zeta|=1}\frac{d\zeta}{2\pi i} P_A 
\begin{bmatrix}
\zeta\Gamma_A+\Gamma_{AB}\Gamma_{BA}+1 & \Gamma_{AB} \\
\Gamma_{BA} & -\zeta\Gamma_B^{-1} + 1
\end{bmatrix}^{-1}
P_A \\
&=\oint_{|\zeta|=1}\frac{d\zeta}{2\pi i} P_A 
\left(1+\zeta\begin{bmatrix}
\Gamma_A-\Gamma_{AB}\Gamma_B^{-1}\Gamma_{BA} & \Gamma_{AB}\Gamma_B^{-1} \\
\Gamma_B^{-1}\Gamma_{BA} & -\Gamma_B^{-1} 
\end{bmatrix}\right)^{-1}
P_A.
\end{split}
\end{equation}
Here we have further assumed $\Gamma_B$ is invertible, and in the last line we have used
\begin{equation}
\begin{bmatrix}
\Gamma_{AB}\Gamma_{BA} +1 & \Gamma_{AB} \\
\Gamma_{BA} & 1
\end{bmatrix}
=
\begin{bmatrix}
1 & \Gamma_{AB} \\
0 & 1
\end{bmatrix}
\begin{bmatrix}
1 & 0 \\
\Gamma_{BA} & 1
\end{bmatrix},\;\;\;\;
\begin{bmatrix}
1 & \Gamma_{AB} \\
0 & 1
\end{bmatrix}^{-1}
=
\begin{bmatrix}
1 & -\Gamma_{AB} \\
0 & 1
\end{bmatrix},\;\;\;\;
\begin{bmatrix}
1 & -\Gamma_{AB} \\
0 & 1
\end{bmatrix}P_A=P_A.
\end{equation}
Likewise, we can obtain the diagonal block restricted to $B$ (under the assumption that $\Gamma_A$ is invertible):
\begin{equation}
\begin{split}
P_B \mathcal{P}_{AB}(\Gamma) P_B &=  \oint_{|\zeta|=1}\frac{d\zeta}{2\pi i} P_B 
\begin{bmatrix}
-\zeta\Gamma_A^{-1}+1 & \Gamma_{AB} \\
\Gamma_{BA} & \zeta\Gamma_B + \Gamma_{BA}\Gamma_{AB}+ 1
\end{bmatrix}^{-1}
P_B \\
&=\oint_{|\zeta|=1}\frac{d\zeta}{2\pi i} P_B 
\left(1+\zeta\begin{bmatrix}
 - \Gamma_A^{-1}& \Gamma_A^{-1}\Gamma_{AB} \\
\Gamma_{BA}\Gamma_A^{-1} & \Gamma_B-\Gamma_{BA}\Gamma_A^{-1}\Gamma_{AB}
\end{bmatrix}\right)^{-1}
P_B.
\end{split}
\end{equation}
In contrast, the off-diagonal block reads
\begin{equation}
\begin{split}
P_A \mathcal{P}_{AB}(\Gamma) P_B &= \oint_{|z|=1} \frac{dz}{2\pi i} (z+\Gamma_A- z\Gamma_{AB}(z\Gamma_B - 1)^{-1}\Gamma_{BA})^{-1}\Gamma_{AB}(1-z\Gamma_B)^{-1} \\
&=\oint_{|z|=1} \frac{dz}{2\pi i} (z+\Gamma_A- \Gamma_{AB}\Gamma_B^{-1}\Gamma_{BA}-\Gamma_{AB}\Gamma_B^{-1}(z - \Gamma_B^{-1})^{-1}\Gamma_B^{-1}\Gamma_{BA})^{-1}\Gamma_{AB}\Gamma_B^{-1}(\Gamma_B^{-1}-z)^{-1} \\
&=\oint_{|z|=1} \frac{dz}{2\pi i} P_A
\left( z+\begin{bmatrix}
\Gamma_A- \Gamma_{AB}\Gamma_B^{-1}\Gamma_{BA} & \Gamma_{AB}\Gamma_B^{-1} \\
\Gamma_B^{-1}\Gamma_{BA} &  - \Gamma_B^{-1}
\end{bmatrix}\right)^{-1} 
P_B,
\end{split}
\end{equation}
whose Hermitian conjugation gives the other off-diagonal block:
\begin{equation}
P_A \mathcal{P}_{AB}(\Gamma) P_B =(P_B \mathcal{P}_{AB}(\Gamma) P_A )^\dag.
\end{equation}

\subsection{Universal bound}
It is clear from the matrices in the above block formulas are closely related to
\begin{equation}
\tilde \Gamma_\pm =\begin{bmatrix}
 - \Gamma_A^{-1}& \pm i\Gamma_A^{-1}\Gamma_{AB} \\
\mp i\Gamma_{BA}\Gamma_A^{-1} & \Gamma_B-\Gamma_{BA}\Gamma_A^{-1}\Gamma_{AB}
\end{bmatrix},
\end{equation}
as well as the inverse
\begin{equation}
\tilde\Gamma_\pm^{-1}=
\begin{bmatrix}
-\Gamma_A+ \Gamma_{AB}\Gamma_B^{-1}\Gamma_{BA} & \pm i\Gamma_{AB}\Gamma_B^{-1} \\
\mp i\Gamma_B^{-1}\Gamma_{BA} &   \Gamma_B^{-1}
\end{bmatrix}.
\end{equation}
Indeed, we can rewrite the diagonal blocks as
\begin{equation}
P_A \mathcal{P}_{AB}(\Gamma) P_A = \oint_{|\zeta|=1}\frac{d\zeta}{2\pi i}P_A  (1-\zeta\tilde\Gamma_\pm^{-1})^{-1} P_A,\;\;\;\;
P_B \mathcal{P}_{AB}(\Gamma) P_B = \oint_{|\zeta|=1}\frac{d\zeta}{2\pi i}P_B  (1-\zeta\tilde\Gamma_\pm)^{-1} P_B,
\end{equation}
and the off-diagonal blocks as
\begin{equation}
P_A \mathcal{P}_{AB}(\Gamma) P_B =\pm i \oint_{|\zeta|=1}\frac{d\zeta}{2\pi i}P_A  (z-\tilde\Gamma_\pm^{-1})^{-1} P_B.
\end{equation}
Note that one can choose either branch of $\tilde\Gamma_\pm$ since $\tilde\Gamma_+ = Z_{AB}\tilde\Gamma_- Z_{AB}$ ($Z_{AB}=\matidentity_A\oplus(-\matidentity_B)=Z_{AB}(0)$) and $P_AZ_{AB}=P_A$, $P_BZ_{AB}=-P_B$. Denoting $P^<_\pm$ / $P^>_\pm$ as the projector onto the subspace spanned by all the eigenvectors of $\tilde \Gamma_\pm$ with eigenvalues whose absolute values are smaller / greater than $1$, we have $P^<_\pm + P^>_\pm=1$, $P^{>,<}_\pm = Z_{AB} P^{>,<}_\mp Z_{AB}$ and
\begin{equation}
P_A \mathcal{P}_{AB}(\Gamma) P_A = -P_AP^<_\pm\tilde\Gamma_\pm P_A,\;\;\;\;
P_B \mathcal{P}_{AB}(\Gamma) P_B = -P_BP^>_\pm\tilde\Gamma_\pm^{-1} P_B,\;\;\;\;
P_A \mathcal{P}_{AB}(\Gamma) P_B =\pm i P_A P^>_\pm P_B.
\end{equation}
Note that $\|P^<_\pm\tilde\Gamma_\pm\|\le1$ and $\|P^>_\pm\tilde\Gamma_\pm^{-1}\|\le1$, we have
\begin{equation}
\|\mathcal{P}_{AB}(\Gamma)\|\le\|P_A\mathcal{P}_{AB}(\Gamma)P_A+P_B\mathcal{P}_{AB}(\Gamma)P_B\| + \|P_A\mathcal{P}_{AB}(\Gamma)P_B+P_B\mathcal{P}_{AB}(\Gamma)P_A\|\le2.
\end{equation}
Recalling Eq.~(\ref{dEdt}) and applying H\"older inequality, we end up with the bound on the negativity change rate (Eq.~\eqref{negativity_rate_bound} in the main text):
\begin{equation}
\left|\pdv{\mathcal{E}}{t}\right| \le \left\| \pdv{\Gamma}{t}\right\|_1.
\label{nn}
\end{equation}

We remark that while the derivation above appears to rely on the invertibility of $\Gamma_{A,B}$, we expect the result to remain valid in general. This is because the divergence should never appear in the final expression due to the projection $P^{<,>}_\pm$. We may also understand the singularity in the negativity dynamics from the ambiguity of $P^{<,>}_\pm$ in case that some eigenvalues of $\tilde\Gamma_\pm$ are exactly of norm $1$. Again, even though $\partial_t \mathcal{E}(t)$ becomes discontinuous, the above bound remains valid. Finally, we mention that Eq.~(\ref{nn}) is probably not optimal, i.e., there could be some coefficient smaller than 1 in the right-hand side.

\subsection{Monotonicity under local operations}
On the level of covariance matrix, a general continuous-time Gaussianity-preserving evolution is given by Eq.~\eqref{cov_dynamics}:
\begin{equation}
\pdv{\Gamma}{t}= -i[h,\Gamma] - \{X,\Gamma\} + 2Y,
\label{dG}
\end{equation}
where $h=4H$ is a purely imaginary anti-symmetric matrix, and $X={\rm Re} M$, $Y=i{\rm Im} M$ with $M=2B$ being a positive semi-definite matrix. This time evolution becomes a local operation if
\begin{equation}
h=h_A\oplus h_B,\;\;\;\;
X=X_A\oplus X_B,\;\;\;\;
Y=Y_A\oplus Y_B,
\label{hXY}
%P_A h P_B = P_A X P_B = P_A Y P_B =0.
\end{equation}
i.e., only block-diagonal components can be nonzero in these matrices. Substituting Eq.~(\ref{dG}) into Eq.~(\ref{dEdt}), we will encounter terms like
\begin{equation}
\mathcal{P}_{AB}(\Gamma)\Gamma = 1 - \int^\pi_{-\pi} \frac{dk}{2\pi} (1+Z_{AB}(k) \Gamma)^{-1},\;\;\;\;
\Gamma\mathcal{P}_{AB}(\Gamma) = 1 - \int^\pi_{-\pi} \frac{dk}{2\pi} (1+ \Gamma Z_{AB}(k))^{-1}.
\end{equation}
Introducing
\begin{equation}
\mathcal{L}_{AB}(\Gamma)= \int^\pi_{-\pi} \frac{dk}{2\pi} (1+Z_{AB}(k) \Gamma)^{-1},\;\;\;\;
\mathcal{R}_{AB}(\Gamma)=\int^\pi_{-\pi} \frac{dk}{2\pi} (1+ \Gamma Z_{AB}(k))^{-1}=\mathcal{L}_{AB}(\Gamma)^\dag,
\end{equation}
we can perform similar calculations as for $\mathcal{P}_{AB}(\Gamma)$ previously, obtaining
\begin{equation}
\begin{split}
P_A\mathcal{L}_{AB}(\Gamma)P_A&=\oint_{|\zeta|=1} \frac{d\zeta}{2\pi i}P_A(\zeta - \tilde\Gamma_\pm^{-1})^{-1} P_A = P_A P^>_\pm P_A,\\
P_B\mathcal{L}_{AB}(\Gamma)P_B&=\oint_{|z|=1} \frac{dz}{2\pi i}P_B(z - \tilde\Gamma_\pm)^{-1} P_B = P_B P^<_\pm P_B,\\
P_B\mathcal{L}_{AB}(\Gamma)P_A&=(P_A\mathcal{L}_{AB}(\Gamma)P_B)^\dag
=\pm i\oint_{|\zeta|=1} \frac{d\zeta}{2\pi i} P_B(1-\zeta\tilde\Gamma_\pm)^{-1} P_A
=\mp i P_B P^>_\pm \tilde\Gamma_\pm^{-1}P_A.
\end{split}
\end{equation}
Recalling Eq.~(\ref{hXY}) in the case of local operations, we find that $d\mathcal{E}/dt$ cannot increase:
\begin{equation}
\begin{split}
\pdv{\mathcal{E}}{t} &=- \Tr[X] + \Tr[ P^>_\pm X_A] + \Tr[P^<_\pm X_B] - \Tr[P^<_\pm\tilde\Gamma_\pm Y_A] - \Tr[P^>_\pm\tilde\Gamma_\pm^{-1} Y_B] \\
&= -\Tr [P^<_\pm (X_A +\tilde\Gamma_\pm Y_A)] -\Tr [P^>_\pm (X_B +\tilde\Gamma_\pm^{-1} Y_B)] \\
&=-\left(\sum_{n:|\lambda_n|<1} \langle n_\pm| X_A + \lambda_n Y_A |n_\pm\rangle+\sum_{n:|\lambda_n|>1} \langle n_\pm| X_B + \lambda^{-1}_n Y_B |n_\pm\rangle\right)\le0.
\end{split}
\end{equation}
Here in the last line, $\tilde\Gamma_\pm = \sum_n \lambda_n|n_\pm\rangle\langle n_\pm|$ ($|n_+\rangle=Z_{AB}|n_-\rangle$; again, $d\mathcal{E}/dt$ never diverges and becomes discontinuous if $|\lambda_n|=1$ for some $n$) and we have used
\begin{equation}
X_{A,B} \pm Y_{A,B}\ge0\;\;\;\;\Leftrightarrow\;\;\;\; X_{A,B} \pm \lambda Y_{A,B}\ge0,\;\;\forall\lambda\in[-1,1],
\end{equation}
where the left-hand side (follows from $M=X+Y\ge0$ and $\overline{M}=X-Y\ge0$) is a special case of the right-hand side, while the right-hand side can be obtained from a nonnegative-coefficient linear combination of the left-hand side.

\subsection{Numerical calculations of dynamical area law}

For the numerical demonstration of the dynamical area law in the main text, we calculated the negativity change rate for two initial conditions: a Charge Density Wave (CDW) state and a randomly sampled mixed Gaussian state. For numerical calculations, it is better to use the block diagonal representation rather than the original formula \eqref{negativity_rate_formula} because the numerical integration of $\mathcal{P}_{AB}$ may become numerically unstable due to the existence of singularity. 
For example, the CDW state yields a singularity because their covariance matrix reaches $\pm1$ in its spectrum and the integral in the definition of $\mathcal{P}_{AB}$ \eqref{PAB_def} depends on how we avoid the poles. We can avoid such numerical instability by utilizing the block diagonal representation, which solely relies on exact diagonalization.

However, even with the block diagonal representation, we need to carefully treat the ambiguity arising when some eigenvalues of $\tilde\Gamma_\pm$ are exactly equal to $\pm1$.
Since this ambiguity corresponds to the fact that $\partial_t\mathcal{E}(t)$ is discontinuous at $t=0$, we used $\partial_t \mathcal{E}(t)$ at $t=10^{-8}$ for the plot to approximate $\lim_{t\to +0} \partial_t \mathcal{E}(t)$.

\begin{comment}

We verified the obtained bound using the tight-binding chain with the particle-loss dynamics. The tight-binding Hamiltonian is given in Eq.~\eqref{TB}. In addition, we consider the particle-loss dissipation at every site, which is described by the Lindblad operator $\hat{L}_\mu = \sqrt{\gamma}\hat{f}_\mu$. We assume that the decay rate $\gamma$ is homogeneous. This yields the matrix $X,Y$ which appears in the equation of motion for the covariance matrix \eqref{cov_dynamics} as $X=(\gamma/2)\matidentity_{2N}, Y=-(\gamma/2)\sigma_y^{\oplus N}$.

In the main text, we showed the two typical examples which include (a) the particle-loss dynamics from a maximally entangled initial pure state and (b) dissipative dynamics from an unentangled initial pure state. In both cases, the initial state yields a singularity because their covariance matrix reaches $\pm1$ in its spectrum and the integral in the definition of $\mathcal{P}$ \eqref{PAB_def} depends on how we avoid the poles. Note that this ambiguity corresponds to the fact that $\partial_t\mathcal{E}(t)$ is discontinuous at $t=0$. Therefore, we used $\partial_t \mathcal{E}(t)$ at $t=10^{-4}$ for the plot of the exact value to approximate $\lim_{t\to +0} \partial_t \mathcal{E}(t)$.
\end{comment}

\end{document}